\documentclass[showkeys,showpacs,superscriptaddress,twocolumn,pra]{revtex4-2}
\usepackage{graphicx}
\usepackage{bm}
\usepackage{color}
\usepackage{inputenc}
\usepackage{subfigure}
\usepackage{amssymb}
\usepackage{amsmath}
\usepackage{ifpdf}
\begin{document}

\title{Effects of optical lattices on bright solitons in spin-orbit coupled Bose-Einstein condensates}
\author{Golam Ali Sekh}
\email{skgolamali@gmail.com}
\affiliation{Department of Physics, Kazi Nazrul University, Asansol-713340, India} 
\author{Benoy Talukdar}
\email{binoyt123@rediffmail.com}
\affiliation{Department of Physics, Visva-Bharati University, Santiniketan-731234, India}
\begin{abstract}
The stationary bright solitons that appear in the ground state of the spin-orbit coupled Bose-Einstein condensate (SOC-BEC) exhibit nodes. We consider SOC-BEC in combined linear and  nonlinear optical lattices and study their effects on the matter-wave bright soliton   and  find that the parameters of the nonlinear lattice or atomic scattering length can be judiciously manipulated to have useful control over the nodes of the soliton. It is seen that the soliton with large number of nodes is less stable compared to one having fewer number of nodes. We infer that the synthetic  spin-orbit coupling  induces instability in the ordinary matter-wave soliton.  
\end{abstract}
\maketitle
%
\section{Introduction}
Atoms are electrically neutral. Thus, we cannot have spin-orbit coupling (SOC) in the cold atoms of Bose-Einstein condensates (BEC). In view of this, a synthetic magnetic was used \cite{R1} to produce SOC interaction in a BEC consisting of two hyperfine states of {$^{87}$Rb} coupled by a Raman Laser. The experimental realization of SOC in ultracold neutral atoms generated renewed interest in both theoretical  and experimental studies \cite{R2,R3,R4} on BECs.  In particular, the tunability of the Raman coupling parameters was found to open many doors to make use of spin-orbit coupled BEC (SOC-BEC)  for {simulating} a variety of phenomena in condensed matter physics including quantum Hall effect, topological insulators and the like \cite{R5,R6,R7}.
 
It is well-known that  BECs with  attractive inter-atomic interaction  can produce stable matter-wave solitons which represent self-trapped modes in the system \cite{R8}.  This provides an opportunity to study nonlinear phenomena in BECs by manipulating the strength of the  trapping potential of the condensate as well as interaction in the trapped atoms. In the last few decades such problems in conventional BECs were studied in some detail \cite{R9,R10,R11,R12,R13} replacing the trapping potential  by (i) a linear optical lattice produced by counter-propagating laser beams and, (ii) at the same time periodically modulating the atomic scattering length via an  optically controlled Feshbach  resonance.
Since in the equation for dynamical evolution of BEC, the  scattering length  multiples a term cubic in the wave function, the spatial modulation in (ii) corresponds to a nonlinear optical lattice. Here we envisage  a study for  the dynamics of spin-orbit coupled (SOC) quasi-one dimensional  BEC  in the presence of both linear and nonlinear optical lattices. We are interested to investigate how does a  bright soliton in the  SOC-BEC respond to changes in the parameters of the  lattice potentials.

The stationary soliton corresponding to the ground state of the conventional BEC without SOC is {node less}. This observation is consistent with the prediction of the so-called ‘no-node’ theorem for the ground state of bosonic system \cite{R14}. The soliton in SOC-BEC is fundamentally different from the conventional one since the spin-orbit coupling breaks the Galilean invariance of the system. This lack of invariance was experimentally demonstrated \cite{R3} by studying the dynamics of SOC-BEC loaded in a translating optical lattice. The violation of Galilean invariance has a number of physical consequences. For example, the  soliton  arising from the ground state of SOC-BEC exhibits nodes and  number of such nodes depends sensitively on the values of  the spin-orbit-coupling parameter. Additionally, the shape of the moving soliton exhibits  dramatic changes with increasing velocity \cite{R15}.

In the present paper we shall work within the framework of a mean-field theory of the many-body system, in which the BEC is governed by Gross-Pitaevskii equation (GPE) in the presence  of spin-orbit coupling and use a variational approach in order to study the effect of optical lattices on the structure of the  bright soliton in the quasi-one dimensional  SOC-BEC. Our primary objective is to critically examine if the parameters of the lattice potentials could be used to provide useful control over the number of nodes of the bright soliton and thereby make attempts to restore the Galilean invariance in the spin-orbit coupled BEC. Studies in restoration  of Galilean invariance is of relatively recent origin and have mainly been undertaken for nuclear force problem \cite{R16,R17}. However, in order to achieve some effective control over the nodes of the SOC-BEC bright soliton, we shall  first examine the behavior of the chemical potentials for the two  pseudo-spin states of the condensate.  The plane-wave solution of the GPE for the SOC-BEC leads to two distinct branches \cite{R2} in the  energy-momentum dispersion relation.  The chemical potentials of the pseudo-spin states are directly related to the upper and  lower branches of the relation.   Studies in the effective control of chemical potentials by using parameters  of the SOC-BEC {are} known to induce population imbalance between pseudo-spin states \cite{R18}. It remains an interesting curiosity if such {studies} could also provide some signature for  controlling  the number of nodes of the soliton.  We shall see that this is indeed true.

In section II we introduce the GPE for the order parameter with two pseudo-spin components in order to provide a useful description of the BEC with experimentally realizable SOC.  In studying the topology of the bright soliton in a quasi-one dimensional SOC-BEC, we shall essentially follow the Ritz optimization procedure \cite{R19} based on variational formulation of the pair of equations in the GPE. In this approach the first variation of the variational functional is made to vanish for suitably chosen trial functions. We introduce a sech ansatz for the spinner wave function of the condensate. Understandably, the ansatz used by us corresponds to  a soliton solution of the GPE. We  make use of the so-called average Lagrangian technique \cite{R20} to investigate the effect of optical lattices on the nodes of the soliton. We begin section III by examining the sensitivity of the chemical potentials $\mu_j$  on the parameters of the condensate. An analytic expression for  $\mu_j$  derived in section II shows that the chemical potentials depend linearly on both Rabi frequency and spin-orbit coupling. Further, we show that the parameters of the nonlinear lattice can be judiciously exploited to reduce the minima of the chemical potentials as well as the width of the matter-wave soliton. This observation pave the path for reducing the number of nodes of the soliton and thereby make attempt to restore the Galilean invariance by squeezing the soliton. Finally in section IV  we summarize our outlook on the present work and make some concluding remarks.

\section{Mean-field model}  Due to spin-orbit coupling (SOC) the degenerate ultracold atoms in a Bose-Einstein condensate (BEC) is divided into two pseudo-spin states. {The inter-atomic interaction in the trapped  condensate is weak enough to be treated accurately within a mean field approximation.  The atom-atom  interaction can, however, be altered by using Feshbach resonance.} The mean-field dynamics of BEC with SOC is governed by the Gross-Pitaevskii equation(GPE)  \cite{R2,R21}
\begin{eqnarray}
i\hbar \frac{\partial \chi}{\partial t}=\left[H_{\rm SOC}+V(x)+ H_{\rm int}\right] \chi.
\label{eq1}
\end{eqnarray}
Here  $\chi \equiv (\tilde{\Phi}_1,\tilde{\Phi}_2)$ gives the wave function corresponding to the hyperfine states labelled by $|\uparrow\rangle\equiv |F=1,\,m_f=0\rangle$ and $|\downarrow\rangle \equiv |F=1,\,m_f=-1\rangle$, and $V(x)$ is the external trapping potential. The single particle Hamiltonian  ($H_{\rm SOC}$) is given by
\begin{eqnarray}
H_{\rm SOC}=\frac{p_x^2}{2m}+\frac{\hbar \alpha}{m} p_x \sigma_z+\frac{\hbar \Omega}{2}\sigma_x,
\label{eq2}
\end{eqnarray}
where $\Omega$ stands for Rabi frequency and $\sigma_{x,z}$, the Pauli spin matrices. The interaction Hamiltonian $H_{\rm int}$ is given by
\begin{eqnarray}
 H_{\rm int}=\begin{pmatrix}
\gamma |\tilde{\Phi}_1|^2+\beta|\tilde{\Phi}^2_2| & 0\\
0 & \gamma |\tilde{\Phi}_2|^2+\beta|\tilde{\Phi}_1|^2 
\end{pmatrix}.
\label{eq3}
\end{eqnarray}
Here $\gamma$ and $\beta$  stand for intra- and inter-atomic interactions. Introducing the  scaled quantities  $x\rightarrow x/\sqrt{\hbar/m\Delta}$, $\tilde{\Phi}_j \rightarrow \tilde{\Phi}_j \sqrt[4]{\hbar/m\Delta}$ and $E\rightarrow E/\hbar\,\Delta$ in term of the single photon detuning $\Delta$ we rewrite the GPE for a SOC-BEC as 
\begin{eqnarray}
i \partial_t \tilde{\Phi}_j\!\!\!\!&=&\!\!\!\!\left(\!\!-\frac{1}{2}\partial_x^2\!+\!i(-1)^j\!\kappa \partial_x\!+\!V(x)\!-\!\gamma |\tilde{\Phi}_{j}|^2\!-\!\beta |\tilde{\Phi}_{3-j}|^2\!\!\right)\tilde{\Phi}_j\nonumber\\
&+&\Omega_m\! \tilde{\Phi}_{3-j}+\zeta \,x\sin(\omega t) \,\tilde{\Phi}_j \hspace{0.75cm} j=1,2.
\label{eq4}
\end{eqnarray}
The spinor states ($\tilde{\Phi}_j$) are coupled by two-counter propagating Raman laser beams. The strength of spin-orbit coupling $\kappa \,(=-\alpha \sqrt{\hbar \Delta/m})$ depends on the relative angle of the incident Raman beams. The scaled Rabi frequency  $\Omega_m (=\Omega/\Delta)$ can be varied by changing the parameters of the Raman lasers. In the presence of linear optical lattice of wavelength $\lambda_L$, the external potential is given by  $V(x)=V_0 \cos(2 \pi x/\lambda_L)$. The nonlinear optical {lattice} (NOL) is, however, obtained by modulating  $\gamma$ and $\beta$ periodically in space in the vicinity of Feshbach resonance. It gives  $\gamma=\gamma_0+V_{nl}(x)$ and $\beta=\beta_0+V_{12n}(x)$ with $V_{nl}(x)=\gamma_1\,\cos(2\pi x/\lambda_N)$ and $V_{12n}(x)=\beta_1\,\cos(2\pi x/\lambda_N)$ \cite{R9,R22}. A similar type of  NOL has recently been considered by Wang {et al.} in \cite{R23} who showed that the NOL can protect the stability of vortex line structures in a  two-dimensional SOC-BEC. { Note that in writing Eq.(\ref{eq4}) we considered a periodic modulation (frequency $\omega$) of the linear lattice along the  direction with lattice tilt $\zeta$ \cite{R24}. This modulation introduces, in the equation, a driving term $\zeta \,x\sin(\omega t)$, that plays an important  role in studying BEC dynamics \cite{R24a,R24b,R24c}.}

{The dominant effect of the rapidly oscillating term in Eq.(\ref{eq4}) is to add an oscillating phase factor to the wave function or order parameter $\tilde{\Phi}_j$ such that we could write \cite{R25}}
\begin{eqnarray}
\tilde{\Phi}_j(x,t)&=&\Phi_j(x,t) \exp[-2 i (\zeta\,x/\omega) \sin^2(\omega t/2)].
\label{eq5}
\end{eqnarray}
{Thus, $\Phi_j(x,t)$  may be regarded as a slowly  varying function of time}. On substitution of Eq.(\ref{eq5}) in Eq. (\ref{eq4}) we obtain
\begin{eqnarray}
i\frac{d\Phi_j}{dt}\!\!\!\!&&=-\frac{1}{2}\frac{d^2\Phi_j}{dx^2}+i\left[\frac{\zeta}{w}+(-1)^j k\right]\frac{d\Phi_j}{dx}+V(x)\Phi_j\nonumber\\
&+&\left[\frac{3 \zeta^2}{4 \omega^2}+(-1)^j\frac{k\zeta}{\omega}\right]\Phi_j{-\left(\gamma|\Phi_j|^2+\beta|\Phi_{3-j}|^2\right)\Phi_j}\nonumber\\&+&\Omega_m \Phi_{3-j},\,\,\,j=1,2.
\label{eq5a}
\end{eqnarray}
In writing Eq.(\ref{eq5a}) we make use of  $\langle\cos^2(\omega t/2)\rangle_t=1/2$, $\langle\cos^4(\omega t/2)\rangle_t= 3/8$, $\langle\cos(\omega t/2)\rangle_t= 0$ \cite{R26}. On may think in an alternative way by rewriting $\cos^2(\omega t/2)=\frac{1}{2}\left[1+\cos(\omega t)\right]$ and then neglecting the rapidly oscillating terms. This is the so-called rotating wave approximation \cite{R27a}.

We restate the initial-boundary value problem in Eq.(\ref{eq5a}) in terms  of an action {functional} which in view of 
\begin{eqnarray}
\Phi_j(x,t)=\phi_j(x)\,{e^{-i\mu\,t}} \,\,\, 
\label{eq6}
\end{eqnarray}
gives the  Lagrangian function
\begin{eqnarray}
{L}&=&{ L}_1+{ L}_2+{L}_{12}+{L}_{nl},
\label{eq7}
\end{eqnarray}
where
\begin{eqnarray}
{L}_j\!\!&=&\!\!\!\mu_j |\phi_j|^2 \!\!-\!\frac{1}{2}\left|\frac{d\phi_j}{dx}\right|^2\!\!\!\!+\!\frac{\gamma_0}{2}|\phi_j|^4\!-\!\!V(x)|\phi_j|^2\!\nonumber\\&-&\!\Omega_m\phi^*_{3-j}\phi_j,\\
\label{eq8}
{L}_{12}\!\!\!&=&\!\!\beta_0 |\phi_1|^2|\phi_2|^2\!+i\, \kappa\left(\phi_1\frac{d\phi^*_1}{dx}\! -\phi_2\frac{d\phi^*_2}{dx} \right),\\
\label{eq9}
{L}_{nl}\!\!\!&=&\!\!\left[\frac{\gamma_1}{2} \left(|\phi_1|^4 + |\phi_2|^4 \right)\! +\!\beta_1  |\phi_1|^2 |\phi_2|^2 \right]\!V_N(x).\,\,\,\,\,\,
\label{eq10}
\end{eqnarray}
Here the reduced ($\mu_j$) and original ($\mu$) chemical potentials are related  by 
\begin{eqnarray}
\mu_j&=&\mu-(-1)^j\frac{\kappa \zeta}{\omega}-
\frac{3\zeta^2}{4 \omega^2}\\
 k_j&=&{\frac{\zeta}{\omega}+(-1)^j\,\kappa}
\end{eqnarray}
In writing Eq. (\ref{eq8}) we have taken $k_j \approx (-1)^j\kappa$ since $\zeta/\omega\,<\,1$ in the high frequency limit of the driving field. However, the chemical potentials ($\mu_j$) remain unequal for $\kappa\neq 0$.

In order to implement Ritz optimization procedure, we use
\begin{eqnarray}
\phi_j(x)&=&A_j\exp[2i(-1)^j\pi x/J]\, {\rm sech}(x/a)
\label{eq11}
\end{eqnarray}
as trial solutions for the stationary states. Here $A_1$, $A_2$, and $a$ are the variational parameters. The value of parameter $J$ can be  taken from the optimization of the chemical potential ($\mu_j$). In Eq.(\ref{eq11}), {the trial solutions are normalized such that the total number of atom in the system, $N=N_1+N_2=2 a( A_1^2+A_2^2$)}. The average Lagrangian $\langle {L}\rangle$ obtained by substituting Eq.(\ref{eq11}) in $\int_{-\infty}^{+\infty}\,L \,dx$ is given by
\begin{eqnarray}
\langle {L}\rangle=\langle {L}_1\rangle + \langle { L}_2\rangle+\langle {L}_{12}\rangle+\langle {L}_{nl}\rangle,
\label{eq12}
 \end{eqnarray}
where 
\begin{eqnarray}
\langle {L}_j\rangle &=&-\frac{N_j}{6 a^2}-\frac{2 \pi^2 a }{J}\sqrt{N_1 N_2}\, \Omega _m \text{csch}\left(\frac{2 \pi ^2 a}{J}\right)+\frac{\gamma_0 N_j^2}{6 a}\nonumber \\&-&\frac{\pi ^2 a}{\lambda_L}{ N_j V_0 \text{csch}\left({\pi^2 a}/{\lambda_L}\right)}-\frac{2 \pi^2 N_j}{J^2}+\mu_j N_j,\\
\label{eq13}
\langle {L}_{12}\rangle &=&\frac{\beta_0 N_1 N_2}{3 a}+\frac{2 \pi  \kappa (N_1+N_2)}{J},\\
\label{eq14}
\left\langle{L}_{nl}\right\rangle &=& \frac{\pi ^2 \gamma_1}{6 \lambda_N^3} \left(\pi ^2 a^2+\lambda_N^2\right) \left(N_1^2+N_2^2\right) {\rm csch}\left(\frac{\pi ^2 a}{\lambda_N}\right)\nonumber\\
&+&\frac{\pi ^2 \beta_1}{3 \lambda_N^3}N_1 N_2 \left(\pi ^2 a^2+\lambda_N^2\right) {\rm csch}\left(\frac{\pi ^2 a}{\lambda_N}\right).
\label{eq15}
\end{eqnarray}

In the Ritz optimization procedure  the variational derivative $\delta \left\langle{L}\right\rangle/\delta N_j$ is made to vanish \cite{R20}. This condition allows us to calculate chemical potential
\begin{eqnarray}
\mu_j&&\!\!\!\!\!=\frac{1}{6 a^2}-\frac{\gamma_0 N_j}{3 a}-\frac{\beta_0 N_{3-j}}{3 a}+\frac{2 \pi ^2}{J^2}-\frac{2 \pi \kappa}{J}\nonumber\\
\!\!\!\!&+&\!\frac{\pi ^2 a V_0 }{\lambda_L }\,{\rm csch}\left(\frac{\pi ^2 a}{\lambda_L}\right)\!+\!\frac{2 \pi ^2 a \sqrt{N_{3-j}}\, \Omega _m }{J \sqrt{N_j}}{\rm csch}\left(\frac{2 \pi ^2 a}{J}\right)\,\,\,\nonumber\\
&-&\frac{\pi ^2  \left(\pi ^2 a^2+\lambda_N^2\right) {\rm csch}\left(\frac{\pi ^2 a}{ \lambda_N}\right)}{3 {\lambda_N}^3}\left(\beta_1 N_{3-j}+\gamma_1 N_j\right)
\label{eq17}
\end{eqnarray}
for the $j^{\rm th}$  spin state. Understandably, the existence  of a stable bound state in the system can be checked by finding a negative minimum ($\mu_m$) in the chemical potential function $\mu_j$ \cite{R22}.

\section{Effects of optical lattices}
We have noted that the nonlinear optical lattice can induce an additional periodic patter in the Bose-Einstein condensates by spatial variation of inter-atomic interaction in the vicinity of  Feshbach resonance. This lattice is found to be  very efficient  in controlling static and dynamical properties of matter wave solitons \cite{R9,R22,R23}.
  In order to envisage a systematic study on the effect of optical lattices we first try to understand {the} relative importance of different parameters related to SOC based on the linear dispersion relation, $\omega_{\pm}=(k_p^2/2+V)\pm\sqrt{\kappa^2 k^2_p+\Omega_m^2}$. 
 It is obtained by taking plane wave solution $\tilde{\Phi}_j=\Phi_{0j}\, {\exp}[i(k_p x-\omega t)]$  $(\Phi_{0j} \ll 1)$ of Eq. (\ref{eq4}) with energy  $\omega$ and momentum $k_p$. Clearly, the dispersion relation in the presence of SOC  gives two branches; the shape and separation of these branches depend on the parameters of the SOC. 
 If $\kappa^2 < \Omega_m$ then both branches {have} a single minimum.  For $\kappa^2>\Omega_m$,  the lower branch contains  two local minima while the upper branch contains a single minimum. 
 In both cases the chemical potentials lie  below $-\Omega_m$ in the semi-infinite gap where only the nonlinear mode can propagate \cite{R2}. Therefore, it is an interesting  curiosity to check how the chemical potential of each spin state changes in the presence of optical latices if $N_1\neq N_2$ {keeping $N$ fixed.}

\begin{figure}[h!]
\begin{center}
\includegraphics[width=0.21\textwidth]{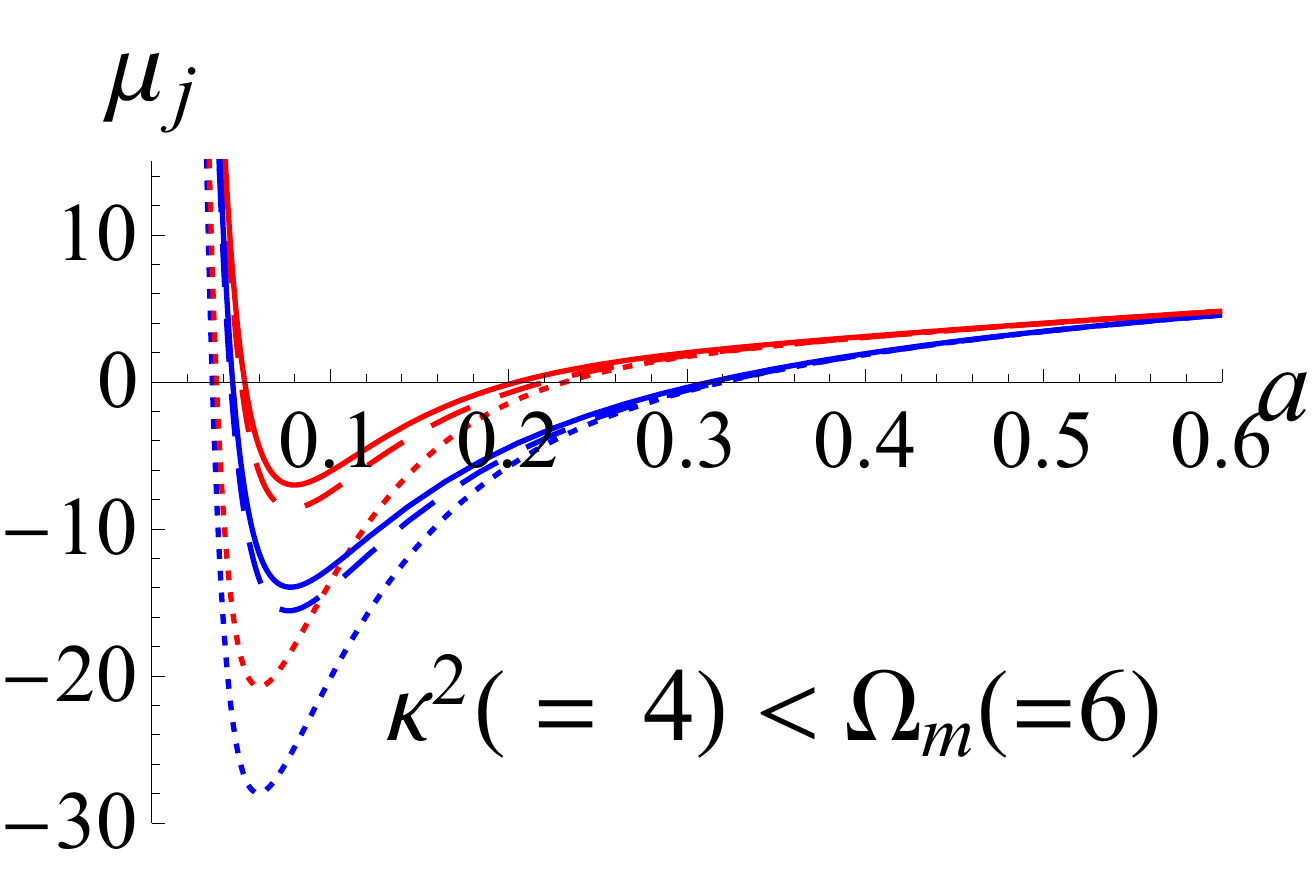}
\includegraphics[width=0.21\textwidth]{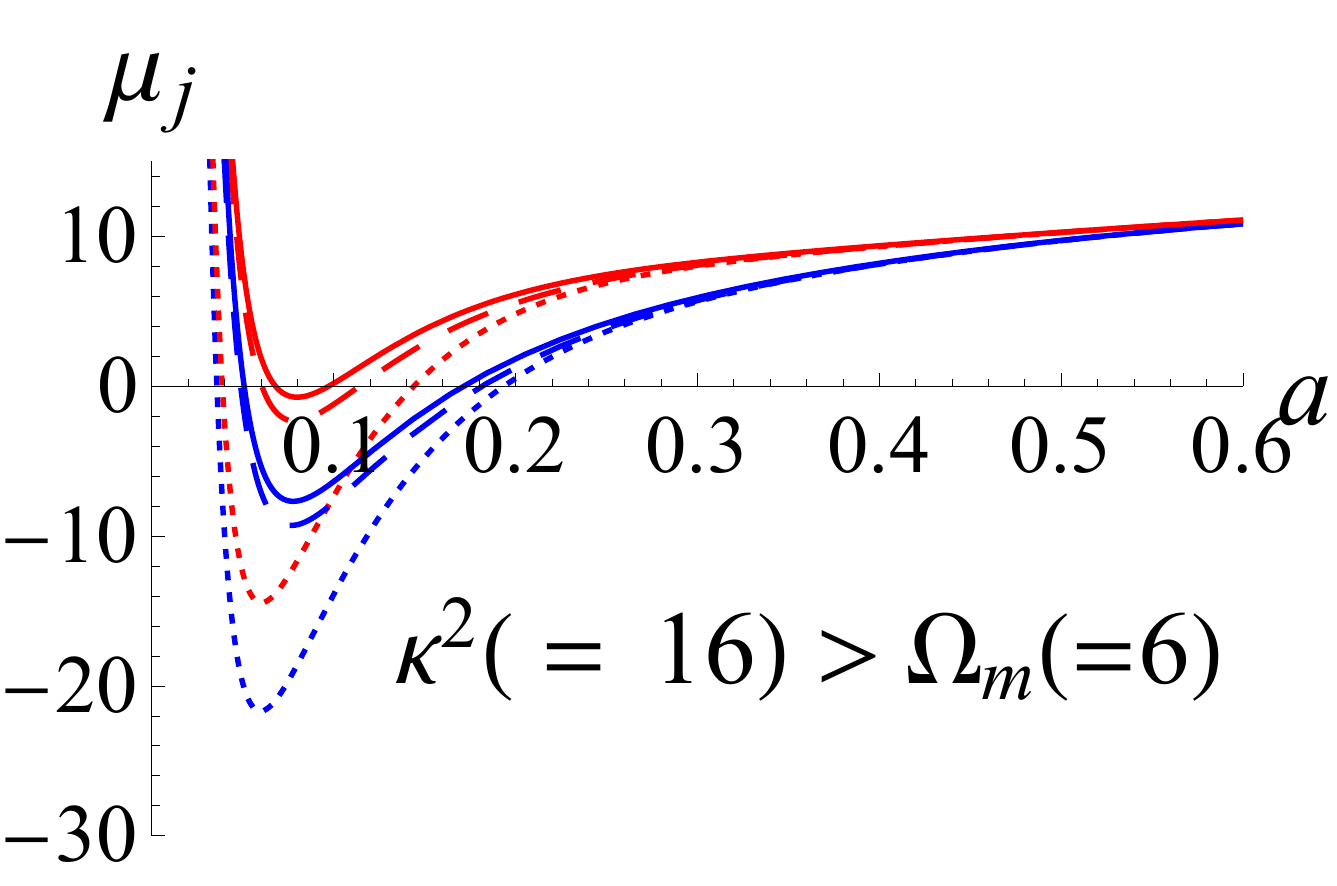}
\vskip -0.1cm
\caption{Variation of chemical potential with width of stationary states for
$\gamma_0=2.0, J=2,$ $\beta_0=2.0,N_1=5.0,N_2=1.5,$  $ \lambda_L =0.65$, $\lambda_N=\lambda_L/2$, 
(i) $\kappa=-2.0$ and $\Omega_m=6$ (left panel) and (ii)$\kappa=-4.0$ and $\Omega_m=6$ (right panel).  Here the red and blue curves represent {results corresponding  to} up and down pseudo spin-states with the plot style : solid (no optical lattices,$V_0=0,\gamma_1=0$ and $\beta_1=0$), dashed (linear optical lattice, $V_0=-2.0$) and dotted (both linear and nonlinear lattices,$V_0=-2.0$, $\gamma_1=0.5$ and $\beta_1=0.5,$ )}
\label{fig1}
\end{center}
\end{figure}

In Fig. \ref{fig1} we plot the chemical potentials ($\mu_j$) as a function of effective width ($a$) considering population imbalanced condition ($N_1\neq N_2$). We see that, for  $\kappa^2 < \Omega_m$, the two branches of chemical potential  deviate appreciably from  each other about their minima ($\mu_m$). The effect of optical lattices causes to decrease the value of  $\mu_m$ (left panel of Fig.\ref{fig1}). More significantly, the action of nonlinear optical lattices reduces the minimum value of chemical potential as well as the effective width of matter-wave solitons. In the case  $\kappa^2 > \Omega_m$, the minimum value of chemical potential ($\mu_m$) is small and negative. Thus, the states are weakly bound.  However, the presence of optical lattices makes the  potential well deeper and thus the states become strongly bound. From this we  infer  that  the NOL can affect stability of matter-wave soliton \cite{R9,R22}. In the following we will see how $\mu_m$ is affected by the variation of different parameters of the system.

\begin{figure}[h!]
\begin{center}
\includegraphics[width=0.22\textwidth]{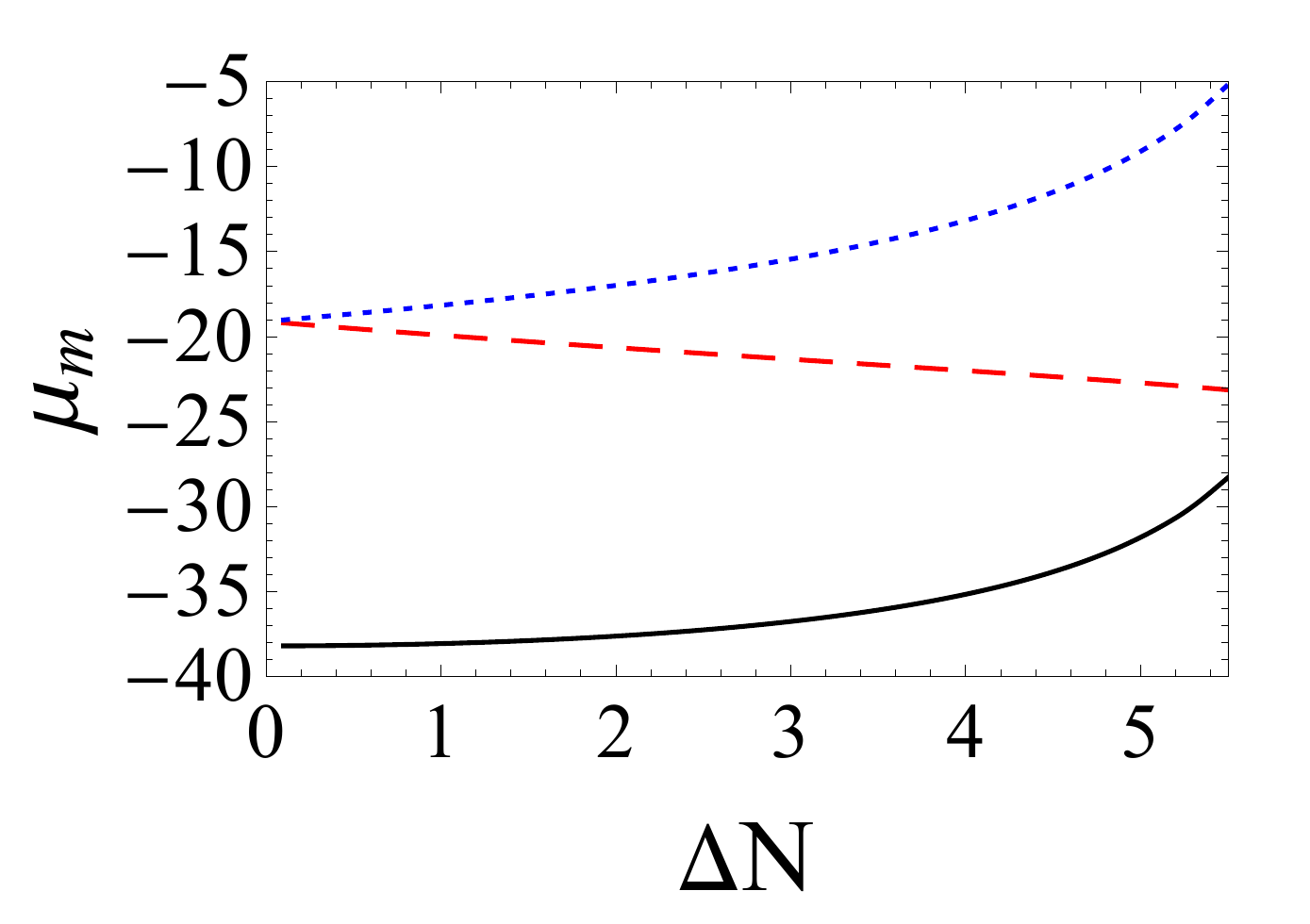}
\includegraphics[width=0.22\textwidth]{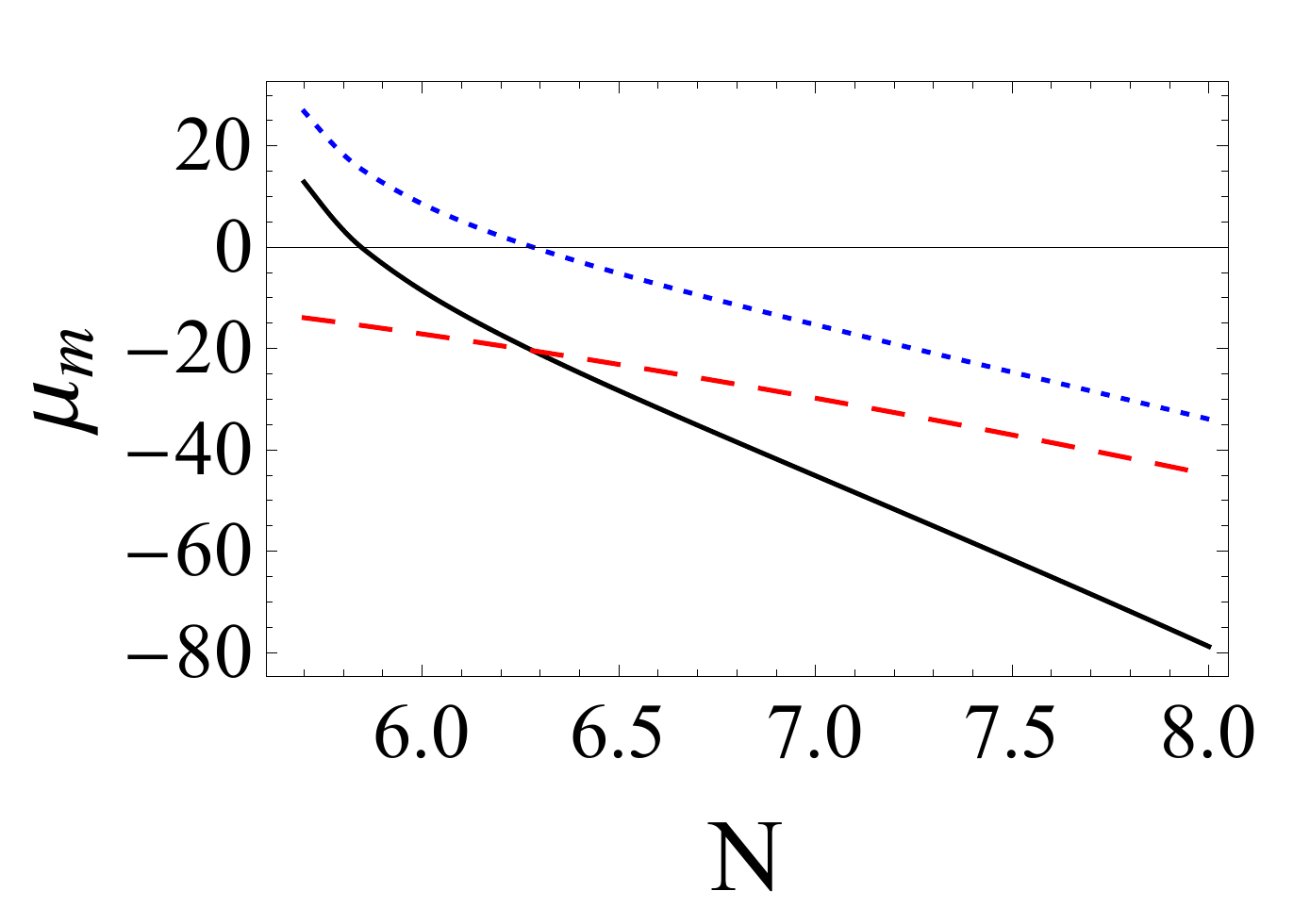}
\vskip -0.35cm
\caption{Left panel: Variation of minimum value of chemical potential $\mu_m$ with $\Delta N$ for $J=2$  and $N=6.5$. Right panel: $\mu_m$ versus $N$ for $\Delta N=3.5$. Red and blue curves represent the first and second components respectively while the black curve  represents the sum of chemical potentials of both  condensates.  Other parameters are kept same with those used in Fig.\ref{fig1}.}
\label{fig2}
\end{center}
\end{figure}
\begin{figure}[h!]
\begin{center}
\includegraphics[width=0.22\textwidth]{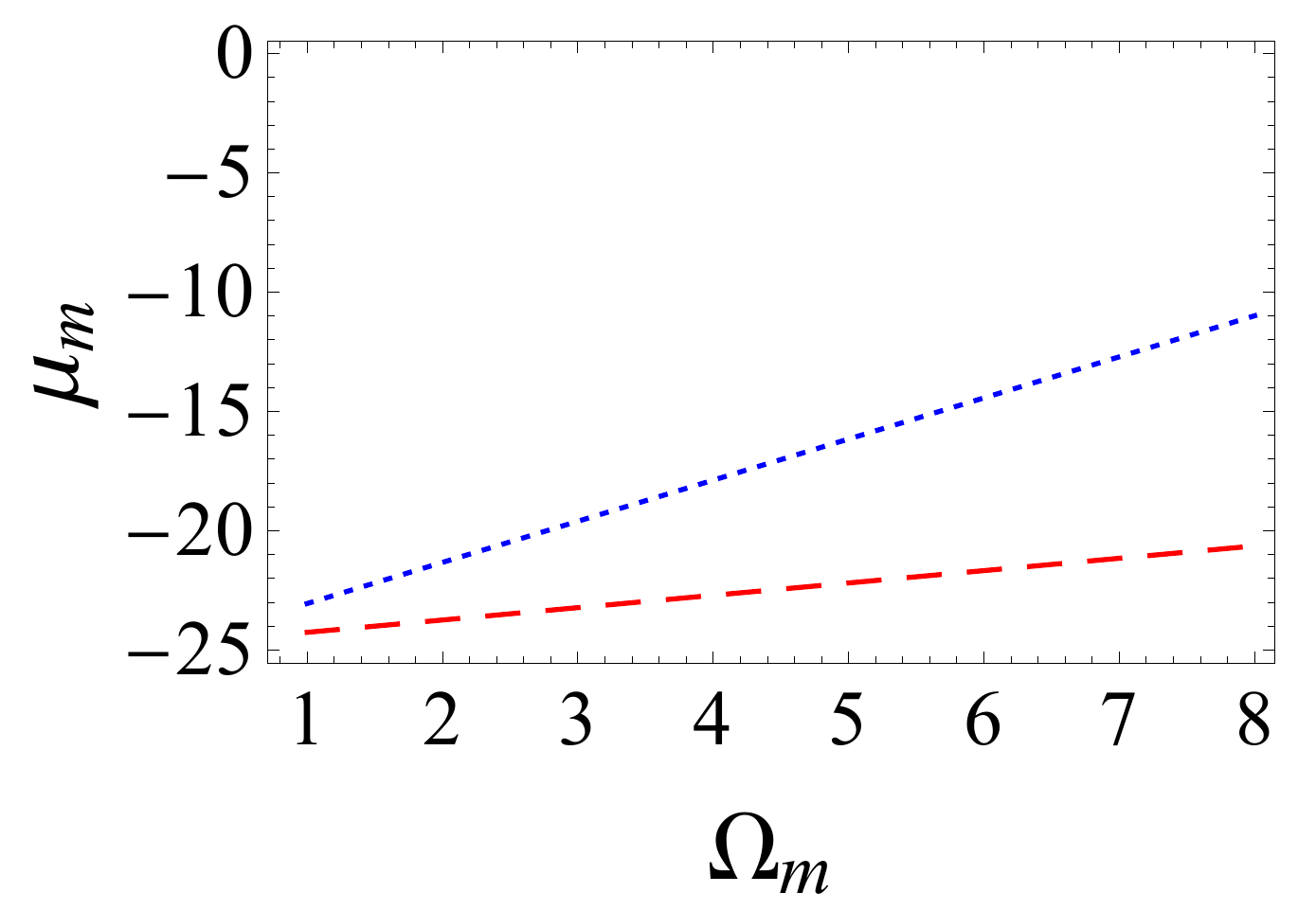}
\includegraphics[width=0.22\textwidth]{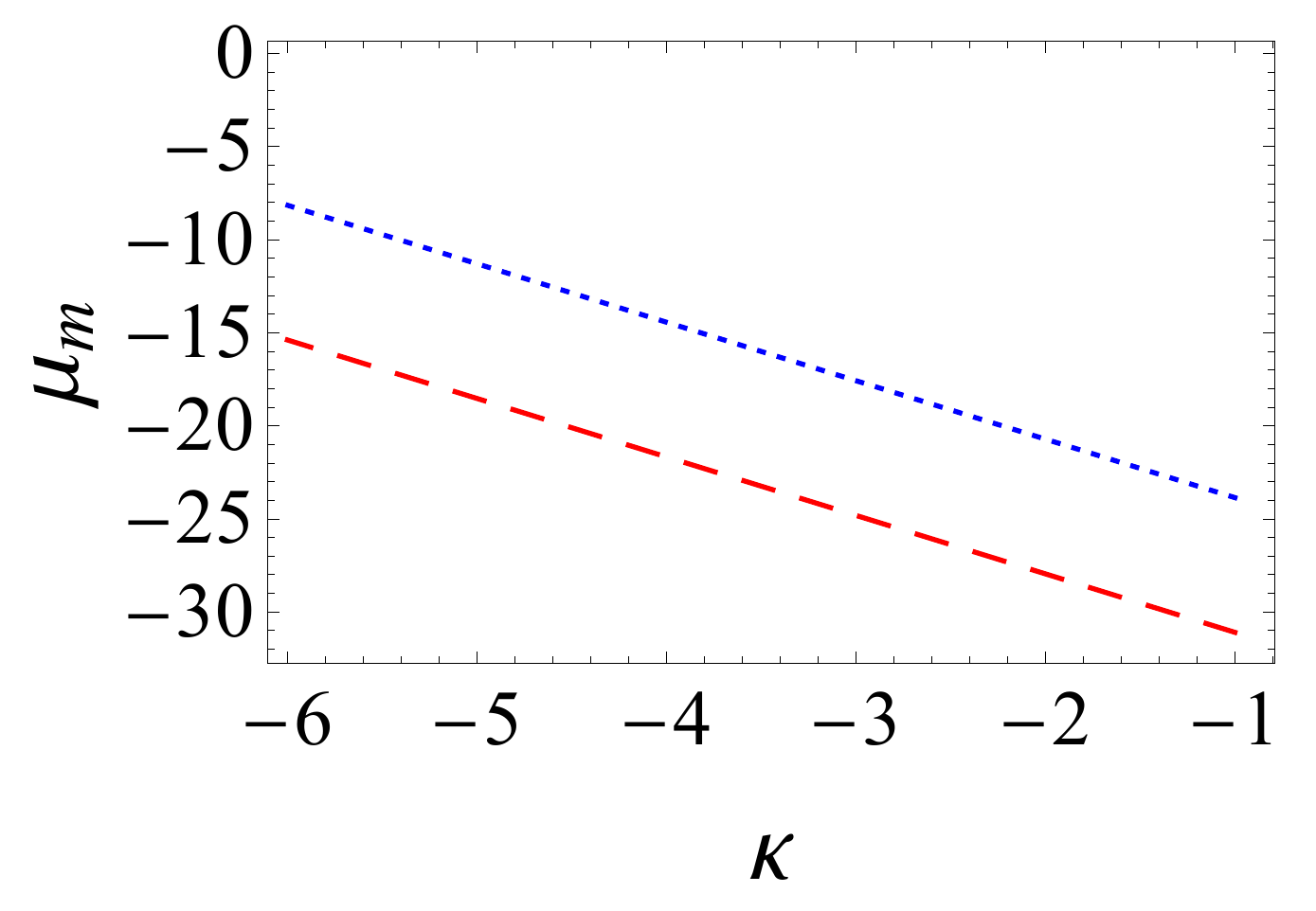}
\vskip -0.35cm
\caption{ Left panel: Variation of $\mu_m$ with $\Omega_m$ for $\kappa=-4$. Right panel: It gives $\mu_m$ with $\kappa$ for $\Omega_m=6$. In both  panels red and blue curves represent the first and second components respectively.  Other parameters are kept same with those used in the right panel of Fig.\ref{fig1}.} 
\label{fig3}
\end{center}
\end{figure}
\begin{figure}[h!]
\begin{center}
\includegraphics[width=0.2\textwidth]{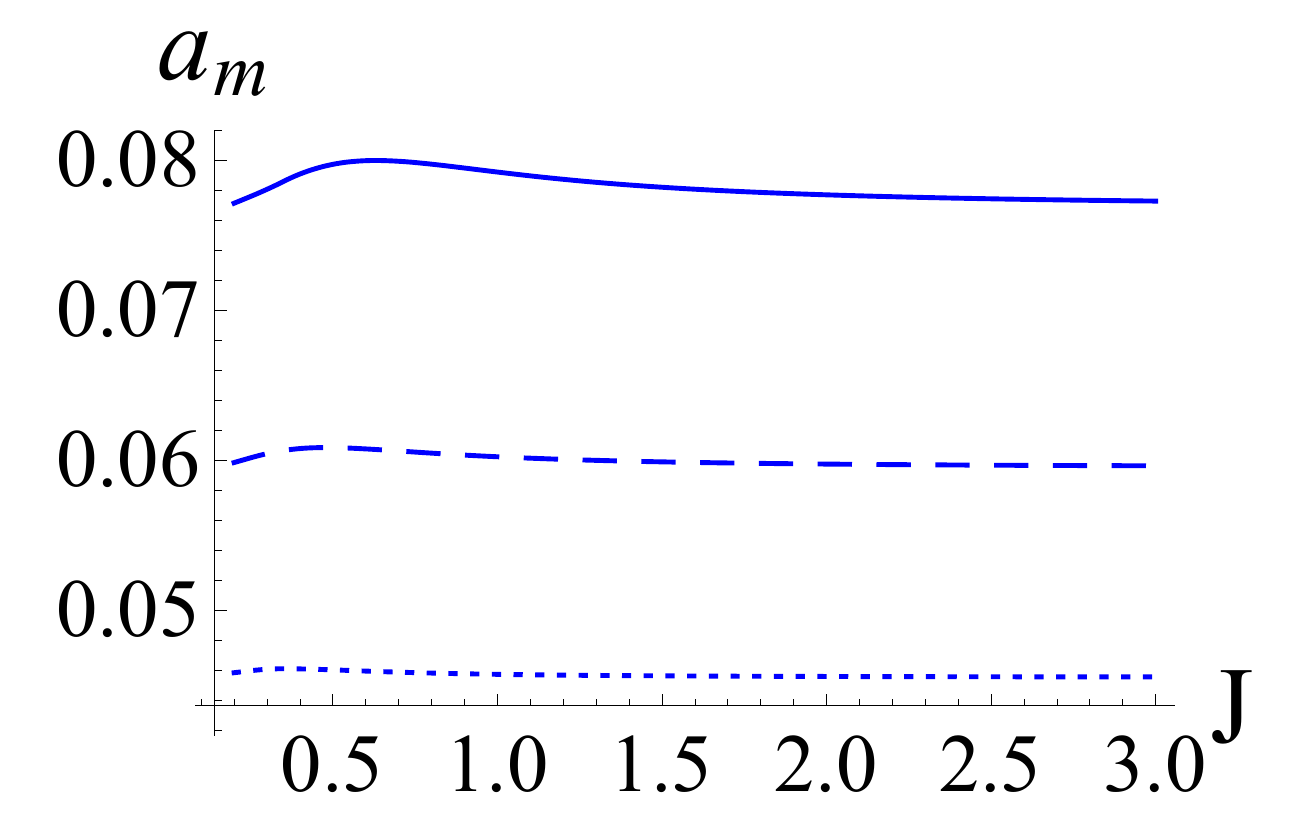}
\hskip -0.185cm
\includegraphics[width=0.2\textwidth]{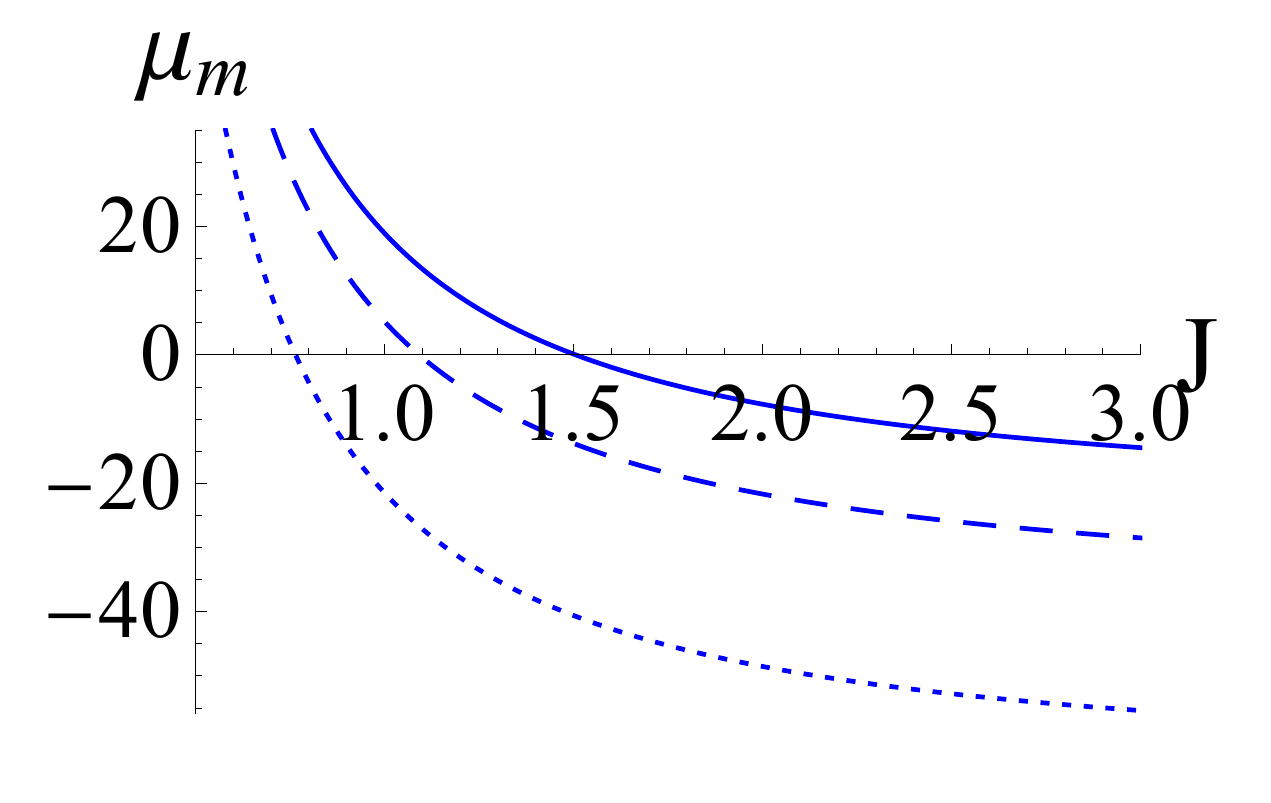}
\vskip -0.1cm
\includegraphics[width=0.2\textwidth]{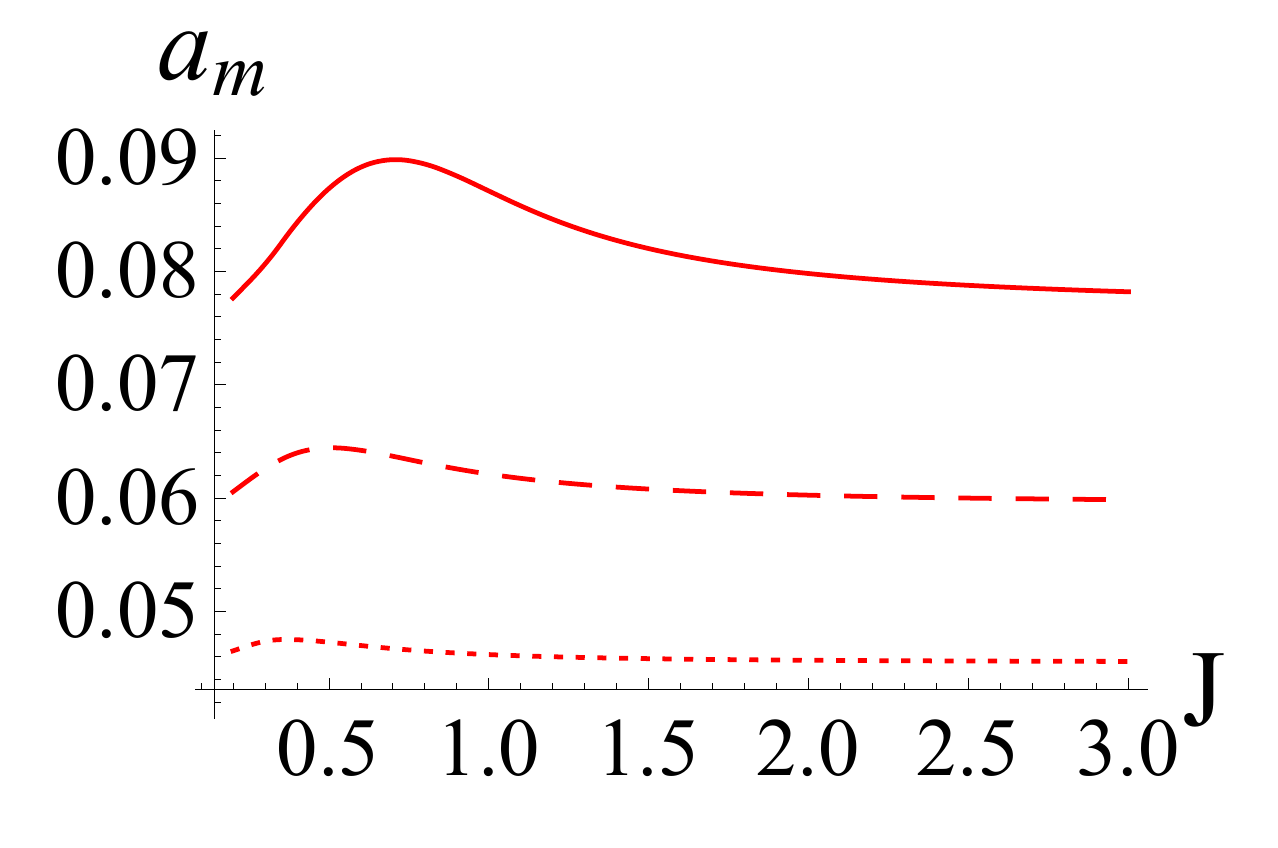}
\hskip -0.185cm
\includegraphics[width=0.2\textwidth]{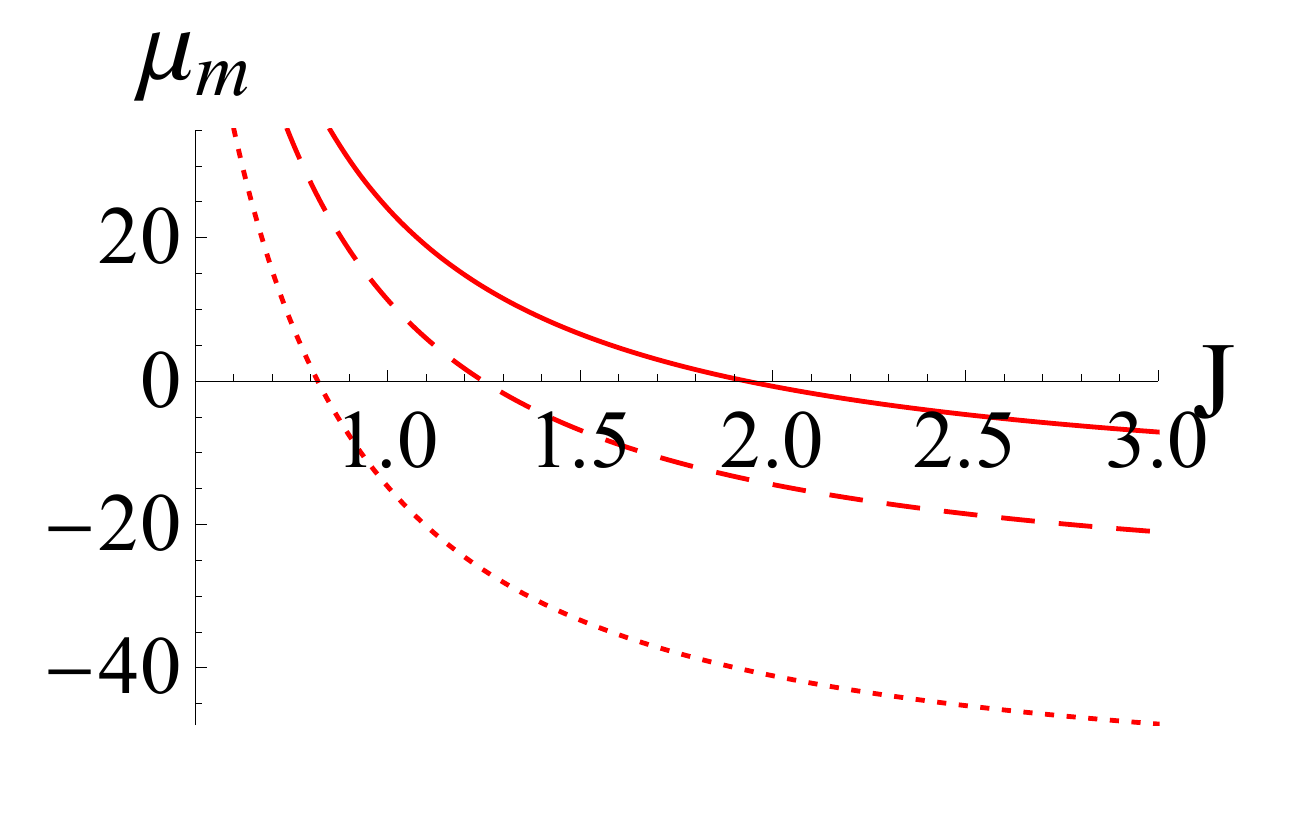}
\vskip -0.1cm
\caption{Variation of $a_m$ and $\mu_m$ as a function of $J$  for (i) $V_0=0,\gamma_1=\beta_1=0$ (solid curve) (ii)$V_0=-2.0,\gamma_1=\beta_1=0.5$ (dashed curve) and (iii) $V_0=-2.0,\gamma_1=\beta_1=1.25$ (dotted curve). Here the red and blue curves represent up and down pseudo spin-states. Other parameters are kept same with those used in the right panel of Fig.\ref{fig1}.}
\label{fig4}
\end{center}
\end{figure}

In the SOC-BEC  population imbalance ($N_1\neq N_2$) can occur due to spontaneous oscillation between  the two pseudo-spin components. From the  viewpoint of thermodynamics, the change in the population between the pseudo-spin components of the BEC  corresponds to going from canonical description to grand-canonical  description of the system. Let $N$ be the total number of atoms in the trap  and $\Delta N$ gives the difference of atoms between two spin states such that $N_1=(N\pm \Delta N)/2$ and $ N_2=(N\mp\Delta N)/2$. Understandably, $\Delta N/N$ defines magnetization ($M$) of the system \cite{R27}. In Fig. \ref{fig2} we plot $\mu_m$  versus $\Delta N$ for a fixed value of $N$. We see that the curves for $\mu_m$  bifurcate for $\Delta N\neq 0$. The separation between the curves  increases gradually as $\Delta N$ increases (left panel of Fig.\ref{fig2}).  Since the value of $N$ is fixed,  the result implies that with the increase of magnetization ($M\propto \Delta N$) in the system the difference of chemical potential  of the two spin states increases. However, for a fixed value of $\Delta N$,  if  we increase $N$ then the values of $\mu_m$ of both  states increase but their difference decrease due to decrease of  magnetization {right panel of Fig.\ref{fig2}}).

\begin{figure}[h!]
\begin{center}
\includegraphics[width=0.19\textwidth]{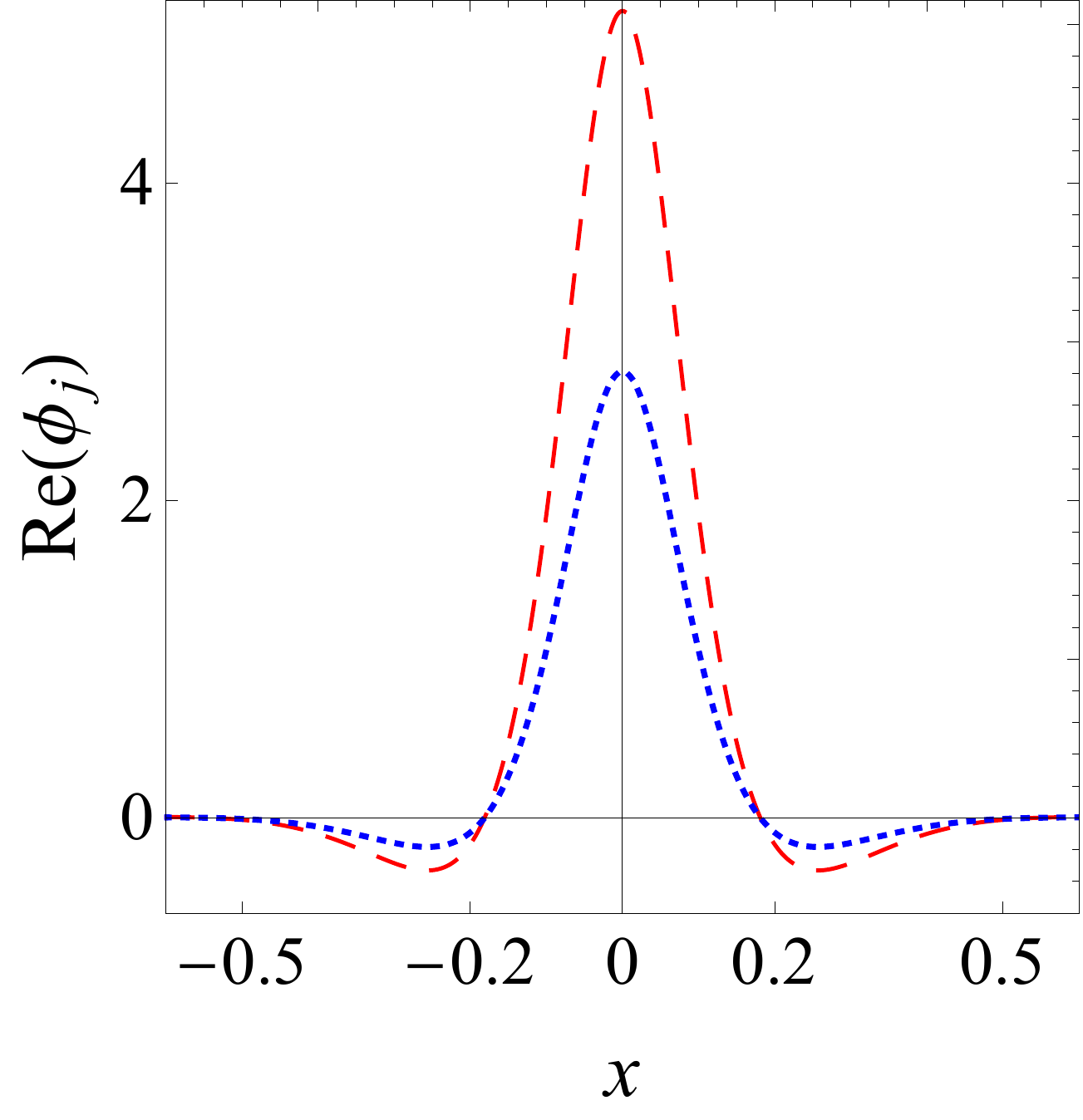}
\includegraphics[width=0.20\textwidth]{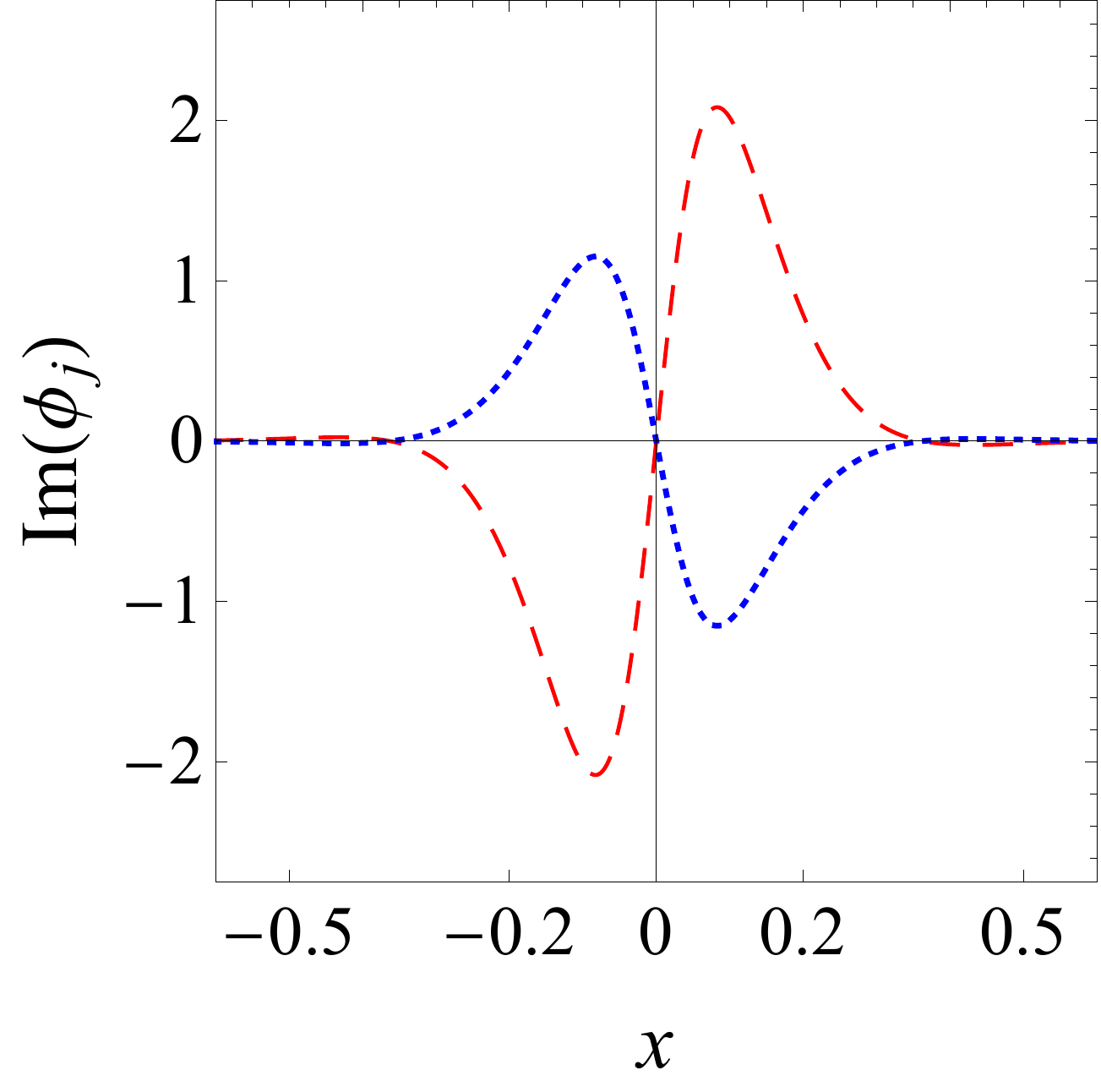}
\includegraphics[width=0.20\textwidth]{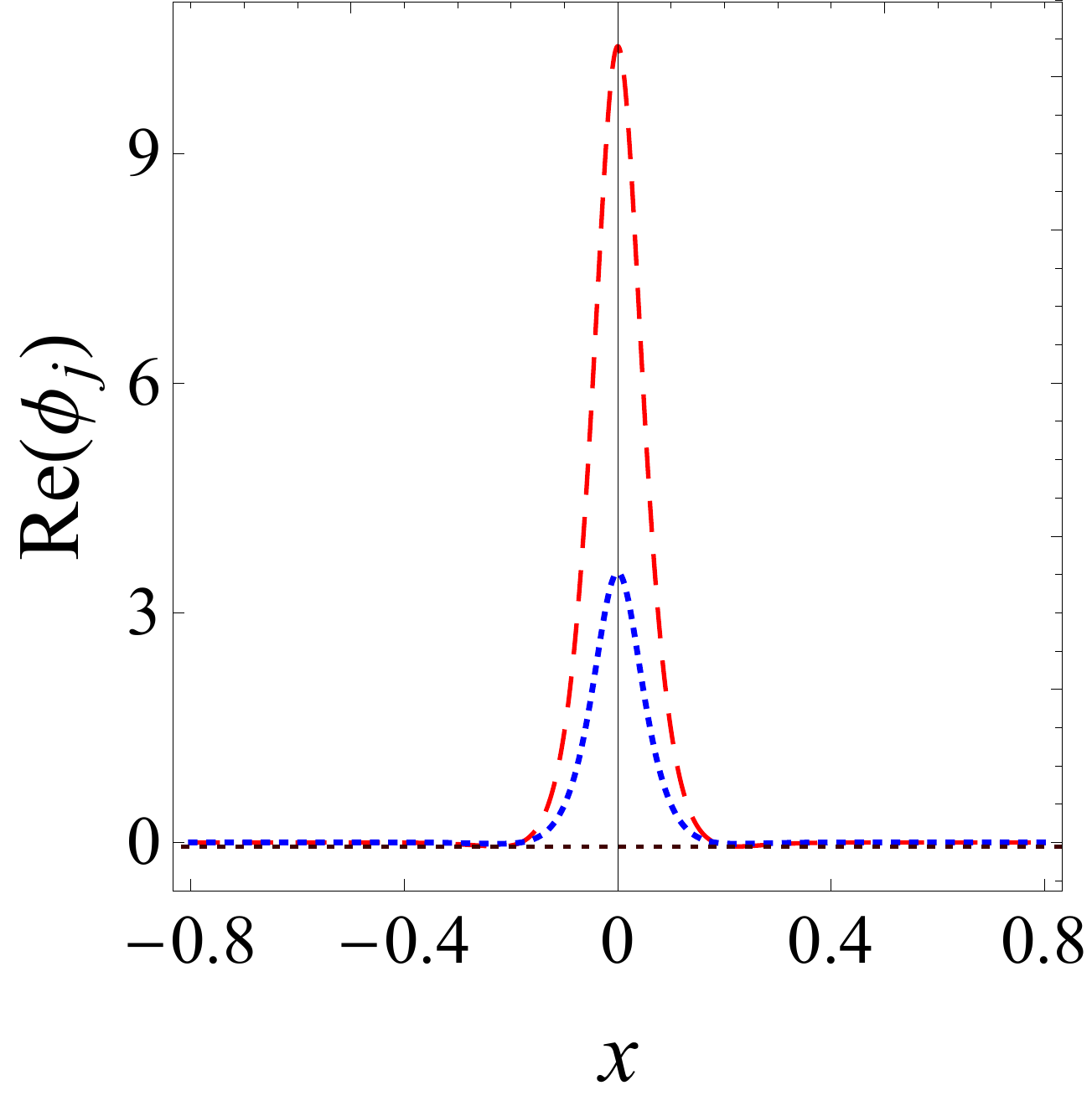}
\includegraphics[width=0.20\textwidth]{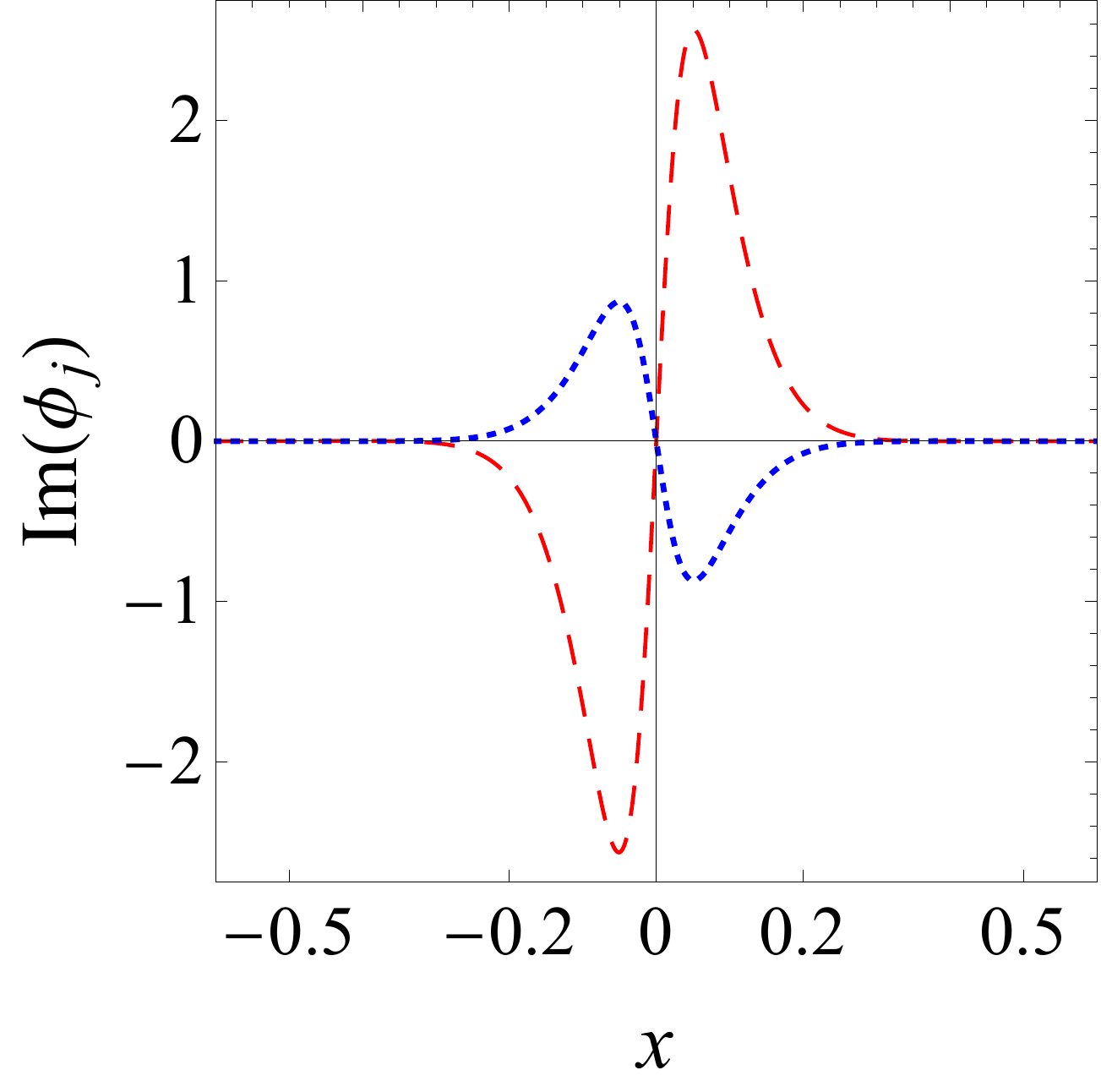}
\vskip -0.35cm
\caption{Spatial variation of real and imaginary parts of $(\phi_j)$ for $J=0.75$. Red and blue {curves represent the order parameters} of the first and second components for  $V_0=-2.0,\gamma_1=\beta_1=0$ (upper panels) while the dashed curve  represents the same for $V_0=-2.0,\gamma_1=\beta_1=1.35$ (lower panels).  Other parameters are kept same with those used in Fig.\ref{fig2}.}
\label{fig5}
\end{center}
\end{figure}

It is relevant to check the  effect of different parameters, namely, Rabi frequency $(\Omega_m)$ and strength of  the SOC $(\kappa)$ for  a fixed value of magnetization ($M=3.5/6.5$). In Fig. \ref{fig3} we display  the  minimum value of $\mu_m$ with the variation of $\Omega_m$ (left panel) and $\kappa$ (right panel).  The figure shows that with the increase of Rabi frequency $\Omega_m$ the fluctuation of number of atoms decreases the minimum value of chemical potential. The initial populations of the spin states are different and thus the effect of $\Omega_m$  on $\mu_m$ is not same. As a result, the difference between the $\mu_m$ values of two spin component increases as $\Omega_m$ becomes larger for a fixed $\kappa$ value. For a fixed value of $\Omega_m$, the values $\mu_m$ shows similar trend with $-\kappa$. However, the difference between the two curves remains almost unchanged. Thus, we see that the effects of $\Omega_m$ and $\kappa$ on $\mu_m$ are distinguishable.

{The matter wave bright soliton in SOC-BEC contains nodes in the ground state. In the proposed trial solution (\ref{eq11}) the parameter  $2\pi/J$ can be regarded roughly as the number of nodes in the soliton with as its overall width. Both these parameters depends sensitively on the strength of the SOC and interaction strength of the nonlinear lattice potential \cite{R15}}. We first ask how does  $a_m$ or $\mu_m$  depend on $J$ for different strength of the NOL. In Fig. \ref{fig4},  we plot $a_m$  and $\mu_m$ versus $J$  for different values of nonlinear lattice strengths. We see that, the value of $a_m$ is small for a small value of $J$.  It first increases with $J$ and  then attains  a maximum value  at $J=J_m$. For $J>J_m$, the value  $a_m$ decreases slowly (left column of {Fig. \ref{fig4})}.
  Its width decreases gradually  with the increase  of lattice strength. Interestingly, the value of $\mu_m$ {is} positive if $J<J_m$. It decreases gradually and becomes negative at a higher values of $J$ (right column of Fig. \ref{fig4}). Thus it would be possible to create weakly bound  or strongly bound state with the variation of $J$.  More specifically,  the soliton with larger number of nodes are either unbound or weakly bound. They become strongly bound for  $J_m \gg 1$.

{We now} focus our attention on the role of NOL to have useful control  over  the number of nodes of the soliton. To that end we first display in Fig.\ref{fig5} the spatial variation of $\phi_j\,\,(j=1,2)$ in the absence of NOL. The red and blue curves give the spin up and spin down components of the condensate wave function. The amplitude of the spin up component is larger than the spin down component. This is true for both the real and imaginary parts of the order parameter. 
{It is clear from the figure that in the absence of NOL both real and imaginary parts of  $\phi_j$ representing the matter wave bright soliton exhibit nodes.}
 We have verified that the number of nodes decreases in the presence of NOL  and nodes disappear if the NOL  is sufficiently strong(bottom panels).  This is due to the fact that the NOL squeezes the matter-wave towards the center of the trap and thus the size of the minimum between two nodes reduces with the  increase of the NOL strength.

{Finally, we  provide some useful checks on the stability of the spin-orbit coupled BEC soliton using the well known Vakhitov-Kolokolov (VK) criterion \cite{R31}.
 In order that we express $N_{1,2}$ in terms of $a$ by solving the equation $\partial \langle L \rangle/\partial a=0$ and get
\begin{eqnarray}
N_1=\frac{2(1+s)+6 a^3(\Gamma_1+\Gamma_2)}{a(\gamma_0+\gamma_0 s^2+2\beta_0 s)+6 a^3\Gamma_3}.
\label{eq18}
\end{eqnarray}
The quantity $s=N_2/N_1$ stands for a control parameter which measures the relative population of two spin components.  In Eq.(\ref{eq18})
\begin{eqnarray}
\Gamma_1\!=\!\frac{ \pi^2 V_0}{\lambda ^2} (s+1)\!\left[\pi ^2 a \coth\! \left(\frac{\pi ^2 a}{\lambda}\right)\!-\!\lambda\,  \text{csch} \left(\frac{\pi ^2 a}{\lambda }\right)\right],\,\,\,\,\,\,\,\,\,\,\,
\label{eq19}
 \end{eqnarray}
 \begin{eqnarray}
 \Gamma_2\!=\!\frac{4 \pi ^2 \Omega _m}{J^2}\sqrt{s} \left[2 \pi ^2 a \coth \left(\frac{2 \pi ^2 a}{J}\right)\!-\!J\right]\! \text{csch}\left(\frac{2 \pi ^2 a}{J}\right)\,\,\,\,\,\,\,\,\,\,\,\,
 \label{eq20}
\end{eqnarray}
 and 
\begin{eqnarray}
\Gamma_3&=&\frac{\pi^4}{6\lambda_N^4} \left[\left(\pi^2 a^2+\lambda_N^2\right) \coth \left(\frac{\pi ^2 a}{\lambda_N}\right)-2 a \lambda_N\right]\times \,\,\,\,\,\,\,\,\,\,\,\,\,\,\,\,\,\,\nonumber\\ &&\text{csch}\left(\frac{\pi ^2 a}{\lambda_N}\right)\left(\gamma_1+\gamma_1 s^2+2 \beta_1 s\right).
\label{eq21}
\end{eqnarray}
We eliminate  $N_j$ in Eq.(\ref{eq17}) by the use of Eq. (\ref{eq18}) and thus get $\mu_j$ as a function $a$. This allows us to write
\begin{eqnarray}
\frac{\partial N_j}{\partial \mu_j}=\frac{\partial N_j/\partial a}{\partial \mu_j/\partial a}.
\label{eq22}
\end{eqnarray}
\begin{figure}[h!]
\begin{center}
\includegraphics[width=0.35\textwidth]{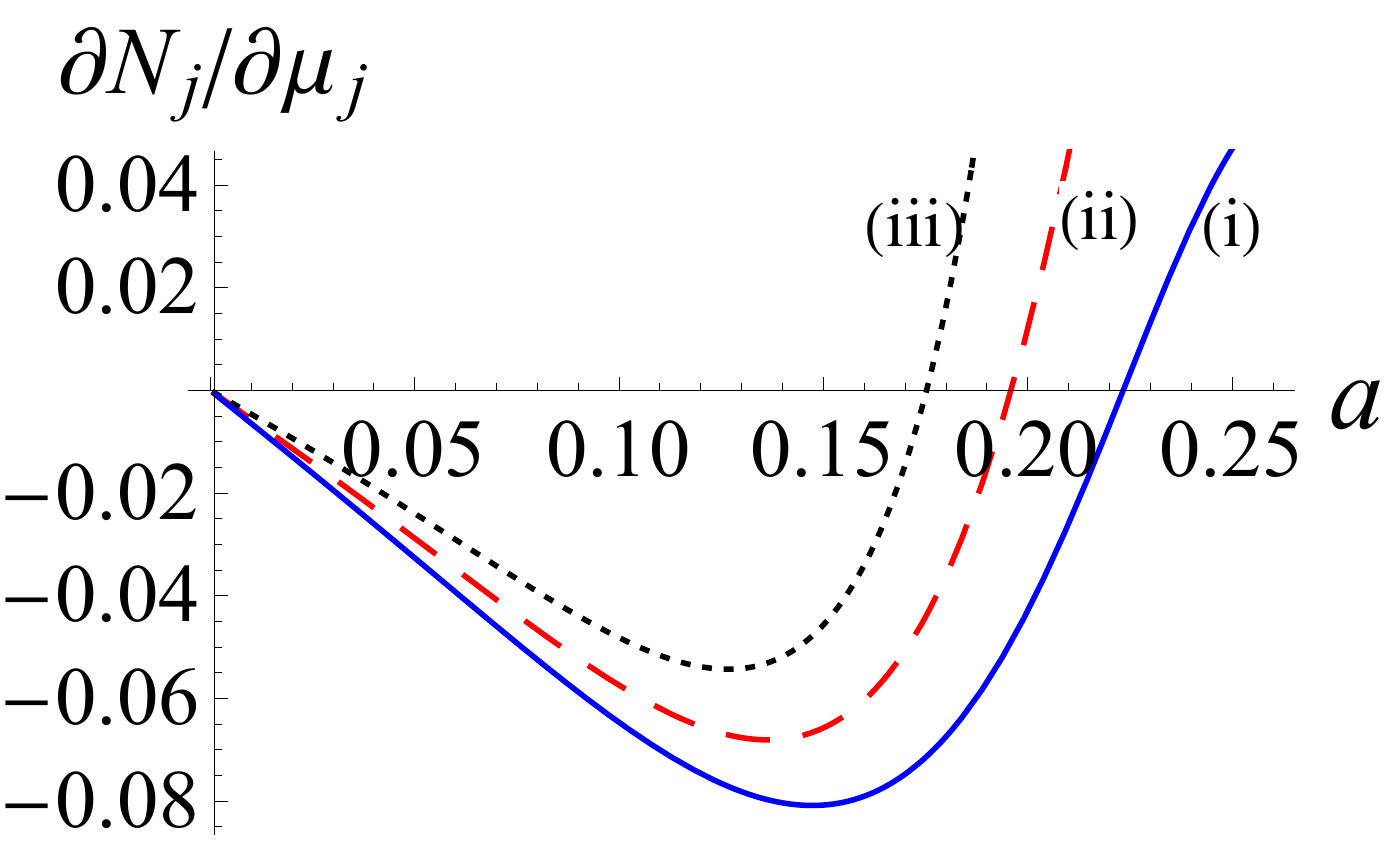}
\vskip -0.35cm
\caption{Variation ${\partial N_j}/{\partial \mu_j}$ versus $a$ with $s=0.30$, $V_0=-2.0$ for different values of $\gamma_1$ and $\beta_1$.  Solid, dashes and dotted give curves   for (i) $\gamma_1=\beta_1=0.4$, (ii) $\gamma_1=\beta_1=0.75$ and (iii)  $\gamma_1=\beta_1=1.35$ respectively. Other parameters are kept same with those used in Fig.\ref{fig1}.}
\label{fig6}
\end{center}
\end{figure}
The VK criterion states that the solitons in SOC-BEC are stable if $\frac{\partial N_j}{\partial \mu_j}\,<\,0$ while they are unstable for  $\frac{\partial N_j}{\partial \mu_j}\,> \,0$. In Fig.\ref{fig6}, we display the variation $\frac{\partial N_j}{\partial \mu_j}$ as a function of $a$ for different strengths  of  nonlinear optical lattice keeping spin-orbit coupling parameter fixed. It is seen  that the matter-wave solitons are linearly stable in the presence of optical lattices {up to} a certain value of its width. More specifically,  width of soliton is squeezed  due to nonlinear optical lattice \cite{R11}. As a result,   amplitude of the soliton increases. On the contrary, the presence of nodes of soliton in SOC-BEC  widens  the effective width of the soliton. Thus, again we find that the effect of nonlinear lattices reduces the nodes of  solitons in spin-orbit coupled Bose-Einstein condensates and leads to the formation of stable fundamental soliton.}

\section{Conclusion}
{Since the seminal work  Lin {et al.} \cite{R1}, studies in the interplay between spin dynamics of Bose -Einstein condensates and nonlinear phenomena have  been regarded as subject of great interest. For example, while examining the effect of random potential  on the dynamics  of SOC-BEC Mardonov {et al,} \cite{R32} found that the spin degrees of freedom of the condensate is influenced  by inter-atomic  interaction  inside the wave packet. Relatively recently, the dynamics of self -attractive SOC-BEC   in a random  potential was studied \cite{R33}  in order to examine how the  soliton motion is affected by Zeeman splitting and self-interaction of the condensate.}

The inclusion of spin-orbit coupling in the atomic Hamiltonian breaks the Galilean invariance of the associated {Schr\"{o}dinger} equation and thus causes splitting of spectral lines. In solids the SOC causes energy bands to split resulting in various physical effects such as the Rashba-Dresselhaus effect or the anomalous Hall effect \cite{R28}. Similarly, in Bose-Einstein condensates the synthetic spin-orbit coupling leads to violation of Galilean invariance of the GPE. As a result there appear nodes in the order parameter of the stationary bright soliton in the ground state of the SOC-BEC . In this work we investigated if the number of such nodes could be controlled by external agencies. To examine this we considered the SOC-BEC loaded in optical lattices and found that the width and number of nodes of the  static soliton  could be controlled by varying the strength of the nonlinear optical lattice or the atom-atom interaction. As the width of the soliton  decreases, the number of nodes becomes fewer. 

The chemical potentials as a function of the width of the soliton form wells. The minima of these wells depends sensitively on the number of nodes of the soliton. The depth  takes large negative values for fewer number of nodes. In contrast, the well becomes shallow or even can take positive values as the number of nodes increases. This implies that the static soliton in the SOC-BEC is less stable than that  found in the ground state of the condensate without spin-orbit coupling. It will be interesting to follow our method to investigate the effect of squeezing on moving solitons of the spin-orbit coupled BEC.

\subsection*{Acknowledgement} One of the authors (GS) would like to  acknowledge the funding from the ``Science Research and Engineering Broad, Govt.of India" through Grant No. CRG/2019/000737.

\end{document}